\documentclass{article}

\usepackage{arxiv}

\usepackage[utf8]{inputenc} % allow utf-8 input
\usepackage[T1]{fontenc}    % use 8-bit T1 fonts
\usepackage{hyperref}       % hyperlinks
\usepackage{url}            % simple URL typesetting
\usepackage{booktabs}       % professional-quality tables
\usepackage{amsfonts}       % blackboard math symbols
\usepackage{nicefrac}       % compact symbols for 1/2, etc.
\usepackage{microtype}      % microtypography
\usepackage{lipsum}
\usepackage{lineno,hyperref}
\usepackage{graphicx, subfigure}
\usepackage{multirow}

\title{The impact of patient clinical information on automated skin cancer detection}

\author{
  Andre G. C. ~Pacheco \\
  Graduate Program in Computer Science, PPGI \\
  Federal University of Espírito Santo, UFES \\
  Av. Fernando Ferrari 514, Vitória-ES, Brazil\\
  \texttt{agcpacheco@inf.ufes.br} \\
   \And
 Renato R. ~Krohling \\
  Production Engineering Department and \\
  Graduate Program in Computer Science, PPGI \\
  Federal University of Espírito Santo, UFES \\
  Av. Fernando Ferrari 514, Vitória-ES, Brazil\\
  \texttt{rkrohling@inf.ufes.br} \\
}

\begin{document}
\maketitle

\begin{abstract}
Skin cancer is one of the most common types of cancer around the world. For this reason, over the past years, different approaches have been proposed to assist detect it. Nonetheless, most of them are based only on dermoscopy images and do not take into account the patient clinical information. In this work, first, we present a new dataset that contains clinical images, acquired from smartphones, and patient clinical information of the skin lesions. Next, we introduce a straightforward approach to combine the clinical data and the images using different well-known deep learning models. These models are applied to the presented dataset using only the images and combining them with the patient clinical information. We present a comprehensive study to show the impact of the clinical data on the final predictions. The results obtained by combining both sets of information show a general improvement of around 7\% in the balanced accuracy for all models. In addition, the statistical test indicates significant differences between the models with and without considering both data. The improvement achieved shows the potential of using patient clinical information in skin cancer detection and indicates that this piece of information is important to leverage skin cancer detection systems.

\end{abstract}

% keywords can be removed

\section{Introduction}
The skin cancer occurs when skin cells are damaged, for example, by overexposure to ultraviolet (UV) radiation from the sun \cite{CCA2018}. Although its incidence data are not required to be reported by most cancer registries \cite{CancerStats2019}, the World Health Organization (WHO) estimates that one in every three cancers diagnosed is a skin cancer \cite{WHO2019}. In countries such as USA, Canada, and Australia, the number of people diagnosed with skin cancer has been increasing at a fairly constant rate over the past decades \cite{CCA2018, CCSsACoC2014, ACS2019}. In Brazil, according to the Brazilian Cancer Institute (INCA), the skin cancer accounts for 33\% of all cancer diagnoses in the country. This is the highest diagnosis rate among all kind of cancer and for 2018-2019 it is expected 180 thousand new cases in the whole country \cite{INCA}.

There are three main types of skin cancer: basal cell carcinoma (BCC), squamous cell carcinoma (SCC) and melanoma. The melanoma is the rarest type of skin cancer, however, due to the high level of metastasis\footnote{when damaged cells invade other parts of the body via blood vessels and lymph vessels}, it is the most lethal one. On the other hand, BCC and SCC, which is known as non-melanoma skin cancer (NMSC), represent the major skin cancer occurrence. As they rarely metastasize, they have low lethality risk \cite{CCA2018}. In order to diagnose the skin cancer, dermatologists screen the suspicious skin lesion using their experience to diagnose it. Moreover, they also take into account clinical information such as patient's age, wherein the lesion is located, if the lesion bleeds, among others \cite{wolff2017, azulay2015}. These pieces of information are very important for dermatologists, nevertheless, differentiate a skin lesion to skin cancer is still challenging. In this sense, to increase the diagnosis reliability, dermatologists use the dermatoscope, a medical instrument that allows the visualization of the subsurface structures of the skin revealing lesion details in colors and textures \cite{argenziano2001}. 

Kittler \textit{et al.} \cite{kittler2002} and Sinz \textit{et al.} \cite{sinz2017} have shown the positive effect of the dermoscopy technique in the diagnostic accuracy. Nonetheless, they also conclude that this accuracy significantly depends on the dermatologist experience degree, i.e., less experienced examiners do not present improvement using the dermatoscope. This reason alone justifies the need for a computer-aided diagnosis (CAD) systems for skin cancer. Nonetheless, in emerging countries, such as Brazil, there is a strong lack of dermatologists and dermatoscopes in most of its countryside cities. Thereby, a system to assist doctors in the skin cancer diagnosis that does not depend on dermoscopy images is very desired. However, designing such a system is a challenging task.

Over the past decades, different computer-aided diagnosis (CAD) systems have been proposed to tackle skin cancer detection. Pioneers works, such as \cite{umbaugh1993}, \cite{ercal1994} and \cite{green1994}, reported the use of low level handcrafted features to differentiate melanomas and NMSC. Later, different computational approaches have been developed based on ABCD(E) rule, pattern analysis and 7-point checklist, which are common methods used by the dermatologist in order to diagnose the skin cancer \cite{argenziano1998, masood2013}. These approaches mostly use traditional computer vision algorithms to extract various features, such as shape, colour, and texture \cite{celebi2007, wighton2011, maglogiannis2015, barata2014, oliveira2018}, to fed a classifier, for example, a support vector machine (SVM) \cite{scharcanski2013, codella2015}. Two weakness may be found in these approaches. First, the ABCD(E) rule and the 7-point checklist were designed only for pigmented lesions, which means they cannot be used to diagnose BCC nor SCC, for example. Second, the handcrafted features extracted by these methods have limited generalization capability \cite{yu2017}.

Recently, deep learning models have been achieving remarkable results in different medical image analysis tasks \cite{litjens2017, yu2017}. In particular, convolutional neural networks (CNN) have become the standard approach to handle this kind of problem \cite{shin2016, tajbakhsh2016}. Several deep learning models have been proposed for skin cancer detection task. Yu \textit{et al.} \cite{yu2017} presented a very deep CNN and a set of schemes to learn under limited training data. Esteva \textit{et al.} \cite{esteva2017} used a pre-trained GoogleNet CNN architecture \cite{szegedy2016} to train more than 120 thousand images and achieved a dermatologist-level diagnostic. Haenssle \textit{et al.} \cite{haenssle2018} and Brinker \textit{et al.} \cite{brinker2019} also used a deep learning models to compare their performance to dermatologists. In both studies, the models have shown competitive or outperformed the dermatologists. Other efforts have been made using deep learning to detect skin cancer, such as ensemble of models \cite{codella2017, harangi2018}, feature aggregation of different models \cite{yu2019}, among others \cite{han2018, attia2017, nida2019melanoma, deAngelo2019}. Most of these works are based on dermoscopy images, mainly for two reasons: 1) there is an open well-known dataset provided by the International Skin Imaging Collaboration (ISIC) \cite{tschandl2019}; 2) obtaining a dataset of clinical images of skin cancer is a hard task. Developing CAD systems to work with dermoscopic images is important, however, as stated before, emerging countries do not have dermatoscopes available in most of their regions. Thereby, these systems are not feasible for those places. Furthermore, there is a trend in developing CAD systems embedded in smartphones, either for general users or to assist doctors \cite{chao2017, ngoo2018fighting}. Indeed, the use of smartphones to assist in skin cancer detection seems to be very feasible. However, it is necessary to have clinical images rather dermoscopy ones.

Beyond the lack of CAD systems using clinical images, most of these systems do not take into account the patient clinical information, which is an important clue towards a more accurate diagnosis \cite{azulay2015}. In fact, the dermatologists do not trust only on the image screening, they also use the patient clinical information in order to provide a more reliable diagnostic. In this sense, Brinker \textit{et al.} \cite{brinker2018} presented a review for deep learning models applied to skin cancer detection and concluded that an improvement in classification quality could be achieved by adding clinical data in the classification process. Based on this idea, Kharazmi \textit{et al.} \cite{kharazmi2018} proposed a deep learning approach to detect BCC using dermoscopic images and five patients clinical information. The results using the clinical data present an improvement, however, they did not analysis how much it affect in the classification. Moreover, they are able to recognize only one type of skin cancer and do not consider clinical images.

In this work, we present a study to analyze the impact of clinical patient information on deep learning models applied to skin cancer detection. In addition, we present a new skin cancer dataset composed by clinical images and patient information and an approach to aggregate the skin cancer images with their respective clinical information. The main contributions of this work are summarized as follows:

\begin{itemize}
    \item In partnership with the Dermatological Assistance Program (PAD) at the Federal University of Espírito Santo (UFES), which is a nonprofit organization that provides skin lesion treatment for low-income people in Brazil, we developed an smartphone application to collect skin lesions images and patient clinical information. From this software, we built a new dataset composed of clinical images and patient clinical information. Since the process of collection of this kind of data is hard, we intend to make this dataset available for research purpose. As far as we know, there is no public skin cancer dataset that contains clinical image and patient clinical information available in the literature.
    
    \item We use well-known deep learning models to develop an approach to aggregate the clinical images and the patients clinical information. This approach introduces a straightforward mechanism to control the contribution of each source of data.
    
    \item We present a comprehensive study to show the impact of the patients clinical data in the skin cancer detection. We analyze the model results with and without using the patient clinical information in order to show the advantages of use this data and how it impacts in the final diagnosis.
\end{itemize}

\noindent The rest of this paper is organized as follows: in section 2 we present the methods and data; in section 3 is presented experiments and results obtained; in section 4 we draw some conclusions.

\section{Material and methods}
In this section, first, we describe the details of the collected dataset. Next, we present a data exploration analysis in order to understand the patient clinical feature patterns. In the following, we describe the models we use in the work and our strategy to combine both the clinical images and the patient clinical information.

\subsection{Dataset} \label{sec:dataset}
In order to acquire clinical images and the patients clinical data, we developed a smartphone-based application to be used by doctors and medical students from the Dermatological Assistant Program (PAD) at the Federal University of Espirito Santo (UFES). Through this application, they attach one or more images of the skin lesion\footnote{In general, we name wounds, moles or spots on the skin as skin lesions. After the diagnosis, the skin cancers will be named as so and the remaining ones will be called by skin diseases} as well as the clinical information related to it. In this sense, each sample in this dataset has a clinical diagnosis, an image and eight clinical information: the patient's age,  the part of the body where the lesion is located, if the lesion itches, bleeds or has bled, hurts, has recently increased, has changed its pattern and if it has an elevation. All these pieces of information are based on the same questions that the PAD's dermatologists ask the patients during the appointment. The application also allows tracking all patient's lesion to follow its evolution throughout the time. Evolution is an important feature and it stands for the E in the ABCD(E) rule. Despite its importance, it will take some years until this information become available, since the lesion may take some time to grow up and the patient needs to return to be assisted. Thereby, when the dermatologists ask the patient if the lesion has increased and if it has changed its pattern, they are trying to obtain information about the lesion's evolution. Nonetheless, as these features are obtained by asking the patients, it is important to note they may describe such information with some subjectivity and imprecision. For this reason, this kind of information must be used to support the diagnosis. The main information is still coming from the screening, i.e., the image.

Regarding the region of the body where the skin lesion is located, there are more than 120 anatomical regions used by the dermatologists. Based on the PAD's dermatologists experience, we grouped all the regions in 15 macro regions that are more frequent and have more potential to arise a skin lesion, they are: face, scalp, nose, lips, ears, neck, chest, abdomen, back, arm, forearm, hand, thigh, shin and foot. As skin lesions have a preference for some regions of the body \cite{wolff2017, azulay2015}, it is an important feature to considerate.

We have been collecting this dataset for one year and a half. There are more than fifty types of skin lesions that were collected by our software. However, most of them are rare and contain only a few samples, which makes them very hard to be used in deep learning models. Thereby, for this work, we decided to use the eight most common skin lesion diagnosed at PAD, which are: Basal Cell Carcinoma (BCC), Squamous Cell Carcinoma (SCC), Actinic Keratosis (ACK), Seborrheic Keratosis (SEK), Bowen's disease (BOD), Lentigo Maligna (LEM), Melanoma (MEL) and Nevus (NEV). As the Bowen's disease and Lentigo Maligna are considered SCC in situ and MEL in situ \cite{wolff2017}, respectively, we clustered them together, which results in six skin lesions in the dataset. In Table \ref{tab:samples} is detailed the number of samples for each type of skin lesion present in the dataset. It is important to note this dataset has three skin cancers, BCC, SCC and MEL, and three skin diseases, which may become cancer if not treated, ACK, SEK and NEV. As stated before, the MEL is the most dangerous but rarest, while SCC and BCC are the most common type of skin cancer. The frequency of these samples in the dataset is according to this statement, which makes the PAD dataset imbalanced. In fact, having imbalanced labels is a peculiarity of skin cancer datasets. For instance, the same issue happens with the ISIC dataset \cite{tschandl2019} and we need to find solutions to tackle it.

\begin{table}
\centering
\caption{The number of samples for each type of diagnosis}
\begin{tabular}{cccc}
\hline
\textbf{Clinical diagnosis} & \textbf{N$^{\circ}$ of images} \\ \hline
Actinic Keratosis (ACK)        & 543 \\
Basal Cell Carcinoma (BCC)     & 442 \\
Melanoma (MEL)                 & 67 \\
Nevus (NEV)                    & 196 \\
Squamous Cell Carcinoma (SCC)  & 149 \\
Seborrheic Keratosis (SEK)     & 215 \\  \hline
\textbf{Total} & \textbf{1612} \\ \hline
\end{tabular}
\label{tab:samples}
\end{table}	

\begin{figure}
    \centering
  \subfigure[BCC\label{fig:samples-bcc}]{
       \includegraphics[width=0.15\linewidth]{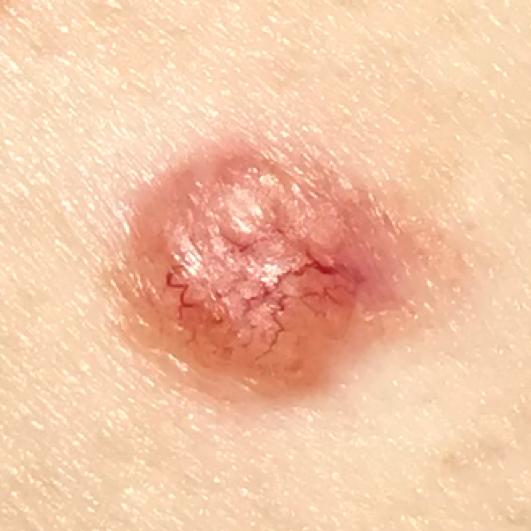}}
    \quad
  \subfigure[SCC\label{fig:samples-scc}]{%
       \includegraphics[width=0.15\linewidth]{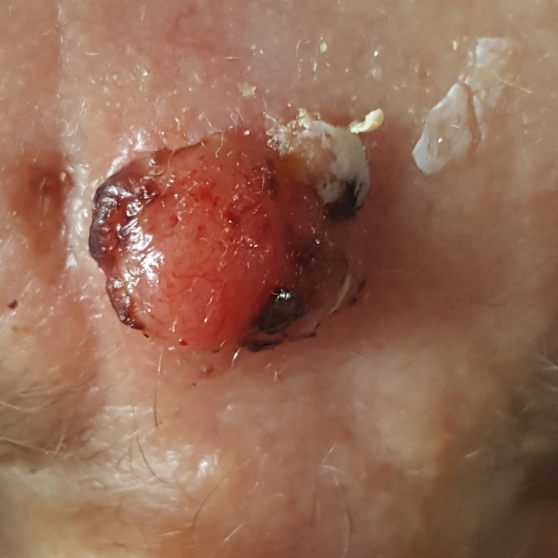}}
    \quad
  \subfigure[ACK\label{fig:samples-ack}]{%
        \includegraphics[width=0.15\linewidth]{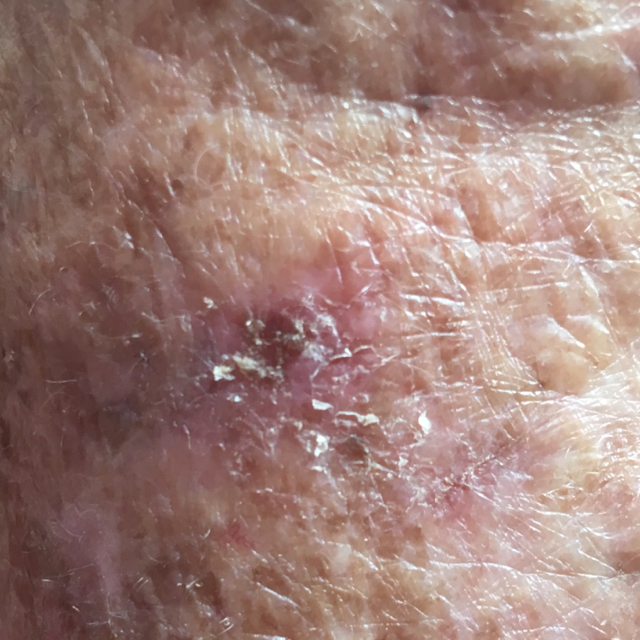}}
    
  \subfigure[MEL\label{fig:samples-mel}]{%
        \includegraphics[width=0.15\linewidth]{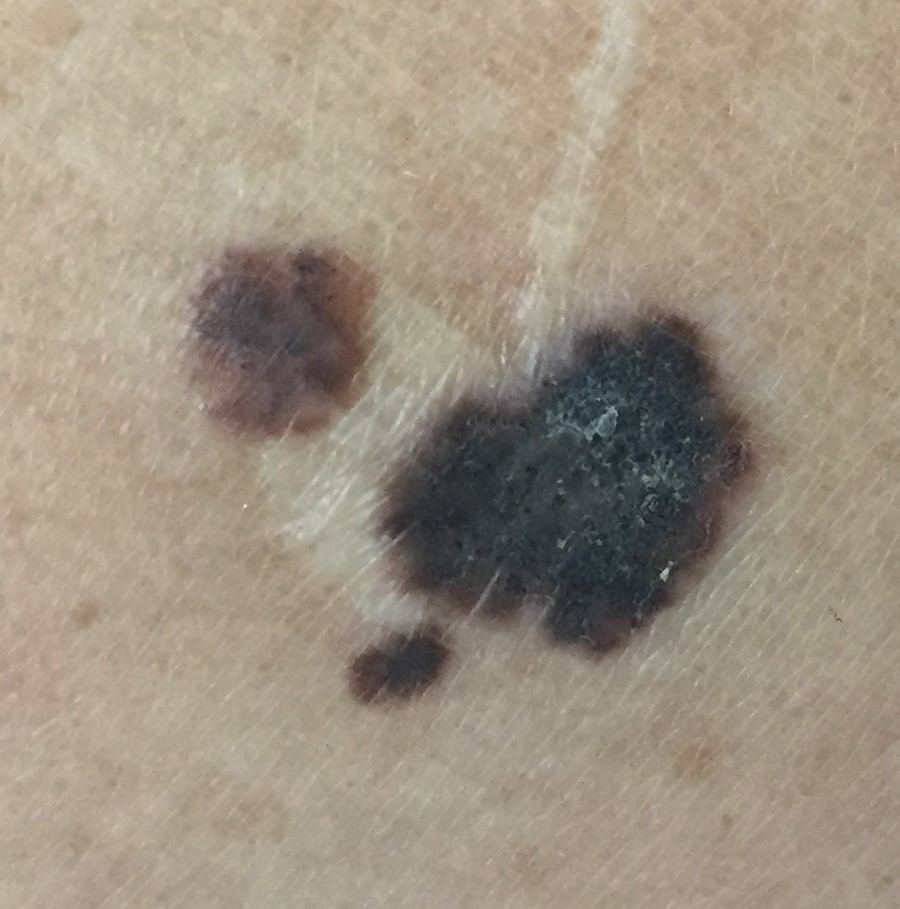}}
    \quad
   \subfigure[NEV\label{fig:samples-nev}]{%
        \includegraphics[width=0.15\linewidth]{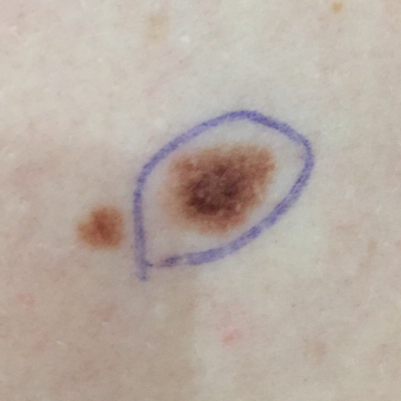}}
    \quad
    \subfigure[SEK\label{fig:samples:f}]{%
        \includegraphics[width=0.15\linewidth]{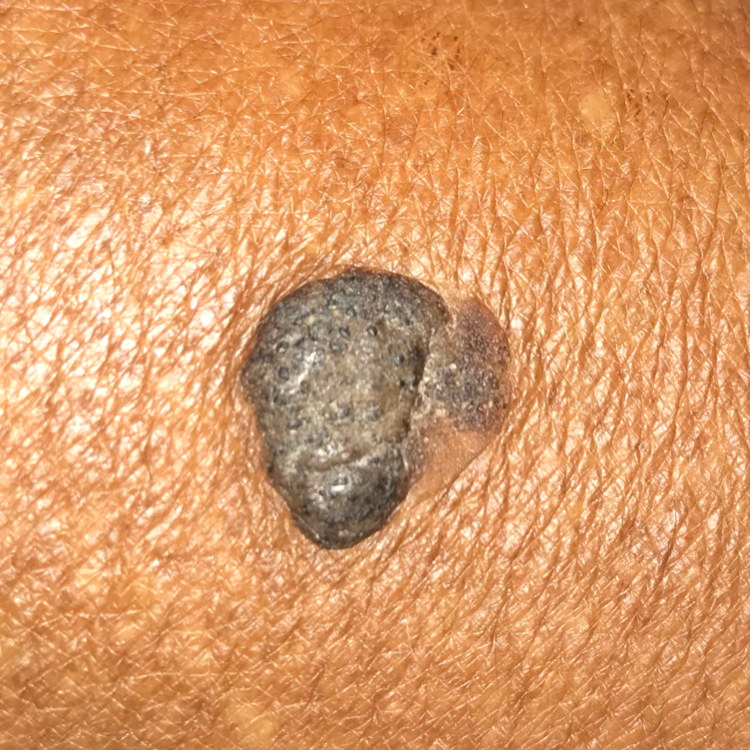}}
        
  \caption{Samples of each type of skin lesion from PAD dataset. SCC, BCC and MEL are skin cancers and NEV, SEK and ACK are skin disease}
  \label{fig:samples} 
\end{figure}

In Figure \ref{fig:samples} is depicted one example for each type of skin lesion present in our dataset. We may note that PAD dataset has three pigmented skin lesions, MEL, NEV and SEK and three non-pigmented ones BCC, SCC and ACK. Lastly, this dataset is available upon request. To the best our knowledge, this is the first public skin cancer dataset composed by images obtained from smartphones and their respectively clinical information.

\subsection{Clinical features analysis} \label{sec:data_analysis}
\begin{figure}
    \centering

    \includegraphics[width=1\linewidth]{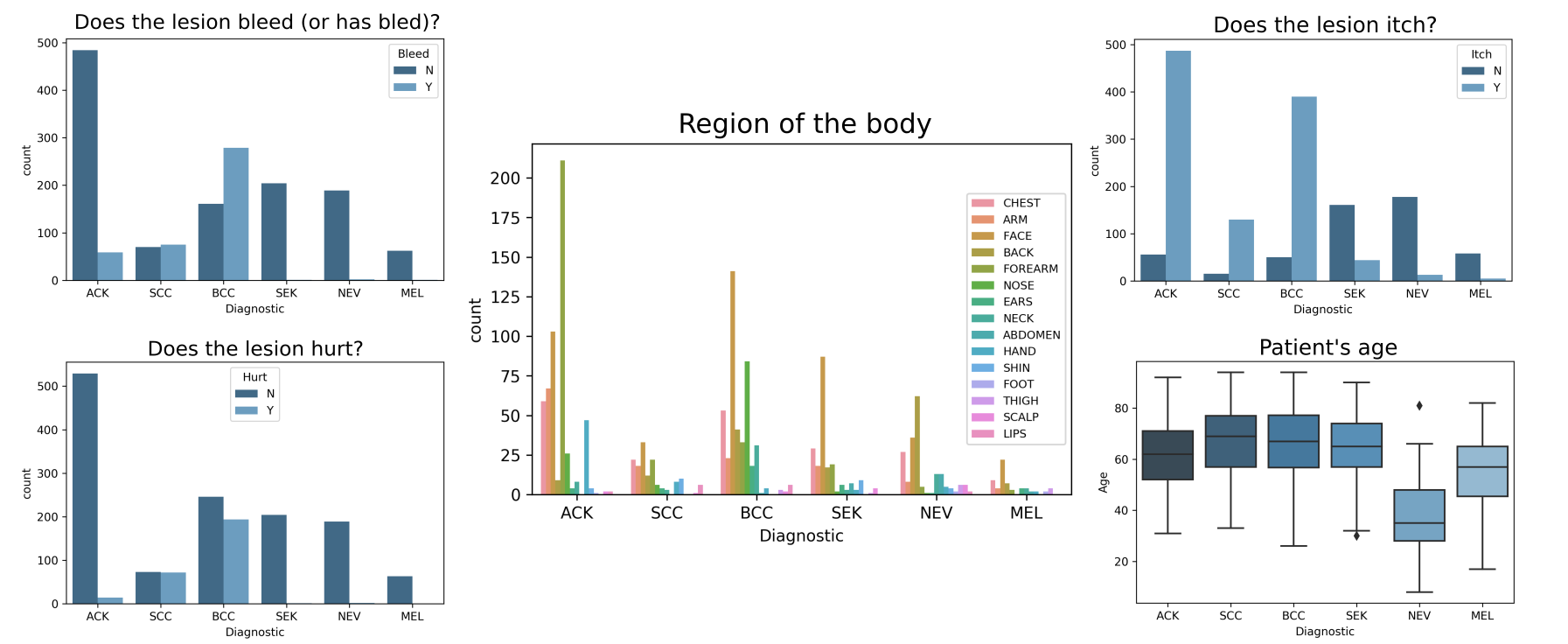}
    
  \caption{Exploratory analysis of the patients clinical features}
  \label{fig:analysis} 
\end{figure}

In order to better understand the influence of the clinical information in the skin cancer detection, we performed a data exploration analysis using the patients clinical features. In Figure \ref{fig:analysis} is shown the plots of the main findings about this set of data. In the left, we observe the plots related to bleeding and pain. As we can see, they are useful to differentiate the pigmented (NEV, MEL, and SEK) from the non-pigmented lesions. Further, we observe that the ACK usually does not hurt. In fact, only SCC and BCC are usually painful lesions. In right side of the Figure \ref{fig:analysis}, we can see a bar plot for itching and a box plot for the patient's age. For itching, again we may observe that in general, the pigmented lesions itch more than the non-pigmented ones. Regarding the patient's age box plot, we can note that the median age for NEV is lower than MEL and SEK. Thereby, this feature is useful to differentiate these lesions. For the non-pigmented lesions, the median for the ACK is slightly lower than SCC and BCC, which lay down in almost the same range. In the center, is presented the plot for the region of the body frequency. Indeed, the lesion present preference for some regions. For instance, The ACK appears more in the forearm, NEV in the back and SCC, BCC, MEL and SEK in the face. We also analyzed the remained collected features. In summary, we observe that MEL and ACK do not have elevation in the skin, which distinguishes them from the other ones; only MEL usually changes its pattern, which is an important feature to detect this type o cancer; and lastly, all lesions, except ACK, usually grow up, but it is hard to find a pattern for this feature.

Based on the presented analysis, we draw the following conclusions: 
\begin{itemize}
    \item It is expected that these features improve the model performance for pigmented and non-pigmented lesions detection.
    
    \item There are punctual features, such as change in the lesion pattern and elevation, that is important for MEL detection.
    
    \item In general, SCC and BCC share the same clinical features values. Both bleed, hurt, itch, have elevation, occur in the same age range and have the same preferred region. Thus, it is expected these features do not help to much for these labels.
\end{itemize}

\subsection{Convolutional Neural Networks} \label{sec:cnn}
The Convolutional Neural Network (CNN) is a special type of a Neural Network (NN) that was developed to learn visual features from images. Nowadays, it is the most successful deep learning methodology to handle the image classification task \cite{rawat2017}. The standard CNN is basically composed of three layers: convolutional, pooling and fully-connected layers. The convolutional layer is the most important and performs the most part of the computation. It consists of kernels, which is composed of weights, that learn visual features from the input images. Each kernel is convolved across the whole image and produces a feature map, which is the output of this layer. The pooling layer is used basically to reduce the feature map size. As a result, it also reduces the number of parameters to be trained in the layers that comes after it, helps to control overfitting and, along with a non-linear activation filter, it adds non-linearity to the network. Lastly, the fully-connected layer is a traditional NN that is connected to the last feature map provided by the previous layer. In summary, the composition of convolutional and pooling layers is known as feature extractor, and the fully-connected layer is the classifier.

Different CNN architectures have been used to tackle skin cancer detection. Successful results have been reported by Steva \textit{et al.} \cite{esteva2017} using GoogleNet \cite{szegedy2015}, Yu \textit{et al.} \cite{yu2019} with ResNet \cite{he2016deep} and Menegola \textit{et al.} \cite{menegola2017} using VGGNet \cite{simonyan2014}. Nonetheless, for medical tasks, obtaining a large amount of data to train a CNN is quite challenging. To overcome this issue, all these works used transfer learning, a well-known technique where a model trained for a given source task is partially reused for a new target task \cite{menegola2017}. Thereby, the models were initialized using the weights from the ImageNet dataset \cite{russakovsky2015} and then fine-tuned using their own dataset.

In this work, our goal is to investigate the impact of clinical information on skin cancer detection. Thus, we decided to use the CNNs mentioned before, i.e, GoogleNet, ResNet50/101 and VGGNet13, but we also included the MobileNet \cite{howard2017}. Each network is briefly described in the following:

\begin{itemize}
    \item \textbf{GoogleNet:} this network introduced the inception module, which is an approach based on several very small convolutions that reduce the number of parameters to be optimized in the training phase \cite{szegedy2015}. It is composed of 9 stacked inception modules that lead to 22 convolutional layers. Also, among these layers, it is performed pooling layers, batch normalization and non-linear activations with ReLU. The feature extractor outputs 1024 image features for the classifier.
    
    \item \textbf{VGGNet-13/19-bn:} this CNN architecture consists of 13/19 layers composed of small convolutional filters \cite{simonyan2014}. It also includes batch normalization, non-linear activations with ReLU and pooling layers after two or three convolutions. The feature extractor outputs 25088 image features for the classifier.
    
    \item \textbf{ResNet-50/101:} this network architecture reformulates the layers as learning residual functions with reference to the layer inputs, instead of learning unreferenced functions \cite{he2016deep}. It is achieved using an approach that skip connections of some layers and applying batch normalization along with non-linearities (ReLU). In this work, we use two ResNet version, one containing 49 convolutional layers and another one containing 100. Both feature extractors return 2048 image features for the classifier.
    
    \item \textbf{MobileNet:} this architecture is based on depth-wise separable convolutions, which is a depth-wise convolution followed by point-wise convolution. This strategy significantly reduces the number of parameters when compared to the network with normal convolutions with the same depth \cite{howard2017}. It also introduces two hyper-parameters to control the model's size. In this work, we use the MobileNet with its full size, which lead to 22 convolutional layers. Similar to the others, it also use ReLU and batch normalization. The feature extractor returns 1024 image features for the classifier.
    
\end{itemize}

\noindent For all these networks, we kept the feature extractor layers and modified the classifier to include the patients clinical information. This is better described in the next subsection.

\subsection{Proposed approach to combine clinical images and clinical information} \label{sec:combine}
In order to consider the patient clinical information into the skin cancer classification, we need to provide a way to combine it with the clinical images. The most common approach to aggregate external features with images is to concatenate the features extracted from the images with the extra ones \cite{ma2016, sun2019multi}. Nonetheless, the issue to use an approach similar to this one for our task is that there are much more image features than patient information, which means we need a feature reducer. In this sense, we propose an approach that uses a NN to reduce the image features based on a mechanism to control the influence of each set of features, i.e., the image features and the patient clinical information.

\begin{figure}
    \centering
    \includegraphics[width=0.7\linewidth]{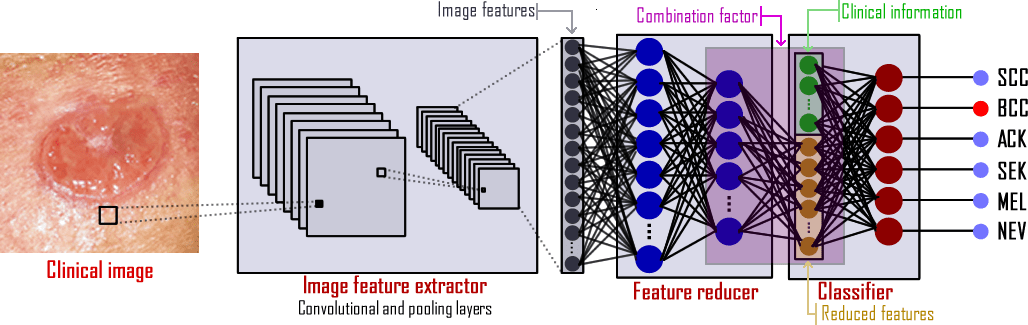}
    \caption{Illustration of the methodology used to combine the image features with the patient clinical information. First, the image features are extracted by the CNN feature extractor. Next, these features are reduced by the feature reducer block and combined with the clinical data. Finally, the combined features are inputted to the classifier that outputs the final diagnostic.}
    \label{fig:propModel}
\end{figure} 

In Figure \ref{fig:propModel} is illustrated the main concept of the proposed aggregation. As we can observe, the CNN's feature extractor is kept, we do not change anything in these layers. Next, the last feature map is flattened and outputted as image features. Next, these features are forwarded to the feature reducer block. This block is composed by a traditional neural network that is trained to work as a non-linear feature reducer. The architecture of the reducer network varies according to the CNN architecture. Further, the number of features outputted by this block is based on the combination factor ($c_f$), which is a mechanism to control how much nodes/image features will be used in the next block. Using this mechanism, we control how much information each source contribute to the concatenation and, consequently, to the classifier. Considering $N_{img}$ and $N_{cli}$ the number of features that comes from the images and clinical information, respectively, to compute the total number of features $T$ to be sent to the classifier, we need to combine both set of features according to the $c_f$:

\begin{equation} \label{eq:total_feat}
T =  \left \lceil c_f N_{img} + (1-c_f) N_{cli} \right \rceil 
\end{equation}

\noindent where $0 \leq  c_f \leq 1$ and $\left \lceil \right \rceil$ is the ceil operator. As we can note from equation \ref{eq:total_feat}, the combination factor $c_f$ controls the amount of feature provided by the image and the clinical information. Nonetheless, in this work, we do not intend to reduce the amount of clinical information since it is already small compared to the image features. Thereby, we set $T = \frac{N_{cli}}{(1-c_f)}$, which $N_{cli}$ is the current value of the clinical features, and we compute $N_{img}$ as follows:

\begin{equation} \label{eq:img_feat_final}
N_{img} = \left \lceil \frac{N_{cli}}{1-c_f} - N_{cli} \right \rceil
\end{equation}

\noindent Based on equation \ref{eq:img_feat_final}, we are keeping the same value of clinical features and varying the value of image features. In section \ref{sec:source_contrib} we present a sensitivity analysis regarding the combination factor.

Since we determined the contribution of each source, both set of features are concatenated and sent to the classifier, as shown in Figure \ref{fig:propModel}. The classifier is another neural network that outputs the probability of a lesion for each diagnostic. The entire model presented in this section is trained by an end-to-end backpropagation. This is the main reason we decide to a neural network as the reducer block, i.e., it can be optimized along with the CNN feature extractor and classifier by including it in the backpropagation computation. We could choose a methodology such as PCA \cite{abdi2010principal}, however, beyond the fact it is linear, the proposed approach is faster and simpler, since the backpropagation is already used to train the image feature extractor blocks. Since the computational cost to train a CNN is high, designing approaches that take advantage of its training process is very desired.

\section{Experiments and results}
In this section, we present the experiments carried out using the presented dataset and the models described in the previous section. First, we describe how we prepared the dataset, next we present the common setup for the experiments, a sensitivity analysis regarding the combination factor and then the comparison between the models using only the clinical images and combining them with the patient clinical features.

\subsection{Data preparation}

\begin{figure}
    \centering
  \subfigure{
       \includegraphics[width=0.15\linewidth]{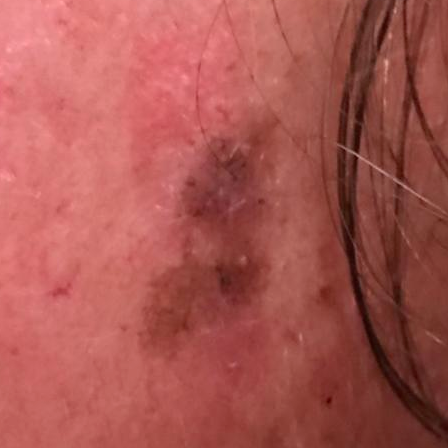}}
    \quad
    \subfigure{%
        \includegraphics[width=0.15\linewidth]{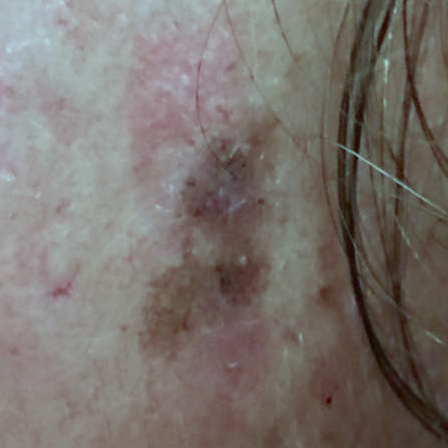}}
    
   \subfigure{%
        \includegraphics[width=0.15\linewidth]{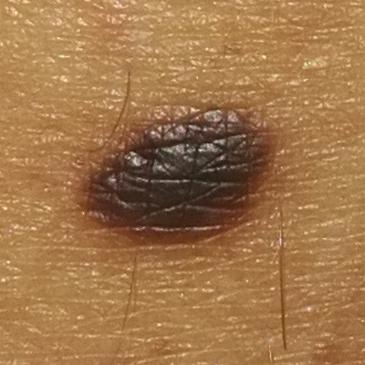}}
    \quad
  \subfigure{%
       \includegraphics[width=0.15\linewidth]{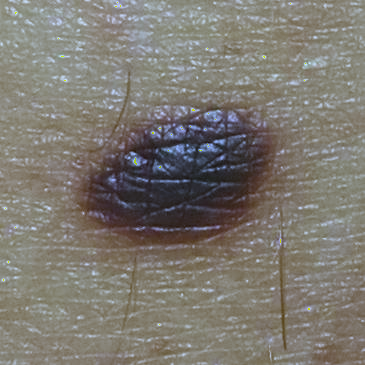}}
       
  \caption{Difference between the original images (left) and the images after the color constancy preprocessing (right)}
  \label{fig:cc} 
\end{figure}

The dataset introduced in section \ref{sec:dataset} presents similar characteristics of any medical dataset. As we may observe in Table \ref{tab:samples}, the amount of data is not large and the dataset is imbalanced. In addition, as it is acquired using smartphones camera, it presents fewer details of the skin lesion, when compared to dermoscopy images, and the camera resolution and illumination affect the images' quality. Vasconceloas and Vasconcelos \cite{vasconcelos2017} presented a work in which they discuss approaches to handle these type of issues in skin cancer datasets. According to them, we can tackle these issues with transfer learning, up/downsampling, data augmentation, and by using an ensemble of models. In addition, Barata \textit{et al.} \cite{barata2014} have shown the benefits of using color constancy algorithms for skin cancer detection. Following their recommendations, we applied the shades of gray method \cite{finlayson2004shades} for all images before the trained phase. The difference between the clinical images with and without the color constancy preprocessing can be seen in the samples depicted in Figure \ref{fig:cc}.

As described in section \ref{sec:cnn}, we use transfer learning for all models. In addition, we applied a strong data augmentation using common image processing operations \cite{perez2018data}. We adjust brightness, contrast, saturation and hue, and we apply horizontal and vertical rotations, translations, re-scale and shear. In addition, we include Gaussian noise and applied blur, since there are some blurred images in the dataset. Lastly, to tackle the labels imbalanced issue, we use a weight loss function based on the labels frequency. Thereby, the weight for a given label $i$ is computed as follow:

\begin{equation}\label{eq:weights}
    w_i = \frac{N}{n_i}
\end{equation}

\noindent where $N$ is the total number of samples in the dataset and $n_i$ is the number of samples of label $i$. We also tried to use upsampling to equalize the number of samples for each class, but it created a high bias in the MEL label, which resulted in a lower performance compared to the weight loss function approach.

Regarding to the clinical features, we applied the one-hot strategy to encode the features represented by strings, i.e., all of them, except the age, which is a integer. Thus, instead of 8 values, after the one-hot encode, the total number of clinical features ($N_{cli})$ is an array of 28 values. As described in equation \ref{eq:img_feat_final}, the size of this array will be used to compute $N_{img}$.

\subsection{Experiments setup} \label{sec:setup}

In order to verify the impact of the patient clinical information, we carried out the experiments considering two scenarios:
\begin{itemize}
    \item Scenario 1: the models consider only the clinical images
    \item Scenario 2: the models combine the clinical images and the patient clinical features
\end{itemize}

\noindent As stated before, for each model, we use its original image feature extractor block. Thus, this block is the same for both scenarios. For scenario 2, we need to define the reducer block described in section \ref{sec:combine}. For all models, except VGGnet, the reducer block is composed by one layer whereas the number of neurons is defined by the combination factor. As the number of image features outputted by VGGnet is more than 25 thousand, the only difference for this model is that we add an intermediate layer containing 1024 neurons. For scenario 1, we tested all models with and without the feature reducer block. Nonetheless, both approaches present the same results. Thereby, for simplicity, we keep the reducer block for this scenario. Also, the reducer block uses the ReLU activation function and dropout rate equal to 0.5.

For both scenarios, the classifiers input neurons are defined by the reducer block and the output for the number of label, which is 6. All models were trained using two phases. First, we freeze the convolutional layers and train only the fully connected one. Next, we reduce the training rate and fine-tune the entire model. For all models, we use the Adam optimizer with learning rate equals 0.0001 for the first phase and 0.00001 for fine-tune. As a loss function we use the weight cross entropy considering the weights computed as described in equation \ref{eq:weights}. The first and second phases are trained for 50 and 100 epochs, respectively. We reduce the learning rate by a rate of 0.1 if the model does not improve for 10 consecutive epochs. We also use early stop if the model does not improve for 15 consecutive epochs. For a fair comparison, both experiments, with and without clinical features, were carried out in the same way and using the same hyperparameters. All procedures were implemented using PyTorch and performed on Nvidia Titan X and RTX 2080 ti. The code is available upon request.

For the following experiments, we use a 5-fold cross validation and present the average and standard deviation for the following metrics: accuracy (ACC), balanced accuracy (BACC), weighted precision (P), weighted recall (R), weighted F1 score (F1) and area under the curve (AUC). Lastly, in order to compare the obtained results by the experiments, we perform the non-parametric Friedman test following by the Wilcoxon test (if applicable), using $p_{value} = 0.05$ and $p_{value} = 0.01$ respectively \cite{derrac2011}.

\subsection{Experiment 1: sensitivity analysis of the combination factor} \label{sec:source_contrib}
In this section, we aim to evaluate the contribution of each source of features in the skin cancer detection. As described in the previous sections, the number of clinical feature ($N_{cli}$) is 28. To vary only the number of image features ($N_{img}$), we keep $N_{cli}$ and vary $c_f$ from 0.5 to 0.9. In Table \ref{tab:cf} is described the the amount of features for each source according to $c_f$. Considering $c_f = 0.7$, it means the reducer block will output 66 features from the image and these features will be concatenated with the 28 clinical features. Thus, 94 features are sent to the classifiers, which 70\% come from the images and 30\% from clinical information. This is how $c_f$ controls the amount of contribution from each source.

\begin{table}
\centering
\small
\caption{The number of features from both sources varying $c_f$}
\begin{tabular}{c|cc|c}
\hline
\textbf{$c_f$}                    & \textbf{$N_{img}$} & \textbf{$N_{cli}$}      & \textbf{Total} \\ \hline
\multicolumn{1}{c|}{\textit{0.5}} & 28                 & \multicolumn{1}{c|}{28} & 56             \\
\multicolumn{1}{c|}{\textit{0.6}} & 42                 & \multicolumn{1}{c|}{28} & 70             \\
\multicolumn{1}{c|}{\textit{0.7}} & 66                 & \multicolumn{1}{c|}{28} & 94            \\
\multicolumn{1}{c|}{\textit{0.8}} & 112                & \multicolumn{1}{c|}{28} & 140            \\
\multicolumn{1}{c|}{\textit{0.9}} & 252                & \multicolumn{1}{c|}{28} & 280            \\ \hline
\end{tabular}

\label{tab:cf} 
\end{table}

In order to test each value of $c_f$ presented in Table \ref{tab:cf}, we applied the ResNet-50 for each folder of our dataset. The result for each metric is detailed in Table \ref{tab:resnet-50-cf}. The Friedman test returned $p_{value} < 0.05$, which means we need to apply the pairwise comparison. Thereby, we applied the Wilcoxon test, that pointed out many differences among the pairs. Based on the test assessment and on the results presented in Table \ref{tab:resnet-50-cf}, we conclude that the best value for $c_f$ is either 0.8 or 0.7. There is no statistical difference between this pair. On the other hand, for 0.5 and 0.9, the results are slightly worse, which is also identified by the test.

\begin{table}
\centering
\small
\caption{The ResNet-50 performance for each value of $c_f$}
\begin{tabular}{c|cccccc}
\hline
\textbf{$c_f$} & \textbf{ACC}      & \textbf{BACC}     & \textbf{P}        & \textbf{R}        & \textbf{F1}       & \textbf{AUC}      \\ \hline
\textit{0.5}   & $0.759 \pm 0.025$ & $0.718 \pm 0.032$ & $0.776 \pm 0.021$ & $0.758 \pm 0.028$ & $0.764 \pm 0.024$ & $0.948 \pm 0.010$ \\
\textit{0.6}   & $0.773 \pm 0.026$ & $0.739 \pm 0.015$ & $0.790 \pm 0.022$ & $0.774 \pm 0.023$ & $0.780 \pm 0.023$ & $0.948 \pm 0.005$ \\
\textit{0.7}   & $0.763 \pm 0.032$ & $0.733 \pm 0.022$ & $0.788 \pm 0.016$ & $0.762 \pm 0.029$ & $0.766 \pm 0.032$ & $0.955 \pm 0.004$ \\
\textit{0.8}   & $0.788 \pm 0.025$ & $0.750 \pm 0.033$ & $0.800 \pm 0.028$ & $0.788 \pm 0.025$ & $0.790 \pm 0.027$ & $0.958 \pm 0.007$ \\
\textit{0.9}   & $0.736 \pm 0.036$ & $0.710 \pm 0.035$ & $0.740 \pm 0.046$ & $0.726 \pm 0.031$ & $0.734 \pm 0.021$ & $0.949 \pm 0.013$ \\ \hline
\end{tabular}
\label{tab:resnet-50-cf}
\end{table}

Based on the presented analysis, we decided to use $c_f = 0.8$ for the next experiment. As discussed before, the clinical features should be used as a support source of information. The main source is still the clinical images. This assumption is in accordance with the results obtained in this section. 

\subsection{Experiment 2: the impact of the clinical features}

In this section, our main goal is to compare the performance of the models for both scenarios described in section \ref{sec:setup}. To do so, according to our previous analysis, we set $c_f = 0.8$ and apply all models described in section \ref{sec:cnn} to the presented dataset. In Table \ref{tab:results-no-feat} and \ref{tab:results-feat} are presented the results considering the 5-folders for scenario 1 and 2, respectively.

\begin{table}[ht]
\centering
\small
\caption{The result for all models in scenario 1, i.e, considering only the clinical images}
\begin{tabular}{c|cccccc}
\hline
\textbf{Model} & \textbf{ACC}      & \textbf{BACC}     & \textbf{P}        & \textbf{R}        & \textbf{F1}       & \textbf{AUC}      \\ \hline
ResNet-50      & $0.671 \pm 0.041$ & $0.649 \pm 0.047$ & $0.720 \pm 0.041$ & $0.670 \pm 0.041$ & $0.678 \pm 0.037$ & $0.927 \pm 0.017$ \\
ResNet101      & $0.691 \pm 0.039$ & $0.651\pm 0.035$  & $0.736\pm 0.028$  & $0.692\pm 0.039$  & $0.700\pm 0.042$  & $0.938\pm 0.008$  \\
GoogleNet      & $0.704\pm 0.024$  & $0.652\pm 0.019$  & $0.714\pm 0.024$  & $0.702\pm 0.025$  & $0.706\pm 0.023$  & $0.927\pm 0.011$  \\
MobileNet      & $0.691\pm 0.024$  & $0.663\pm 0.027$  & $0.720\pm 0.014$  & $0.690\pm 0.025$  & $0.698\pm 0.018$  & $0.932\pm 0.008$  \\
VGGNet-13      & $0.707 \pm 0.028$ & $0.658\pm 0.045$  & $0.734\pm 0.029$  & $0.708\pm 0.028$  & $0.710\pm 0.029$  & $0.932\pm 0.010$  \\
VGGNet-19      & $0.679\pm 0.020$  & $0.628\pm 0.012$  & $0.696\pm 0.020$  & $0.678\pm 0.019$  & $0.680\pm 0.021$  & $0.919\pm 0.009$  \\ \hline
\textbf{AVG}   & $0.690 \pm 0.029$ & $0.650\pm 0.031$  & $0.720\pm 0.026$  & $0.690\pm 0.030$  & $0.695\pm 0.028$  & $0.929\pm 0.011$  \\ \hline
\end{tabular}
\label{tab:results-no-feat}
\end{table}

\begin{table}[ht]
\centering
\small
\caption{The result for all models in scenario 2, i.e, considering both the clinical features and images}
\begin{tabular}{c|cccccc}
\hline
\textbf{Model} & \textbf{ACC}      & \textbf{BACC}     & \textbf{P}       & \textbf{R}       & \textbf{F1}      & \textbf{AUC}     \\ \hline
ResNet-50      & $0.788 \pm 0.025$ & $0.750 \pm 0.033$ & $0.800\pm 0.028$ & $0.788\pm 0.025$ & $0.790\pm 0.027$ & $0.958\pm 0.007$ \\
ResNet101      & $0.757\pm 0.021$  & $0.711\pm 0.019$  & $0.784\pm 0.021$ & $0.756\pm 0.021$ & $0.766\pm 0.022$ & $0.953\pm 0.003$ \\
GoogleNet      & $0.779\pm 0.011$  & $0.714\pm 0.028$  & $0.780\pm 0.011$ & $0.780\pm 0.009$ & $0.778\pm 0.007$ & $0.948\pm 0.008$ \\
MobileNet      & $0.762\pm 0.040$  & $0.717\pm 0.020$  & $0.774\pm 0.031$ & $0.762\pm 0.042$ & $0.762\pm 0.037$ & $0.948\pm 0.013$ \\
VGGNet-13      & $0.746\pm 0.027$  & $0.704\pm 0.007$  & $0.758\pm 0.013$ & $0.748\pm 0.029$ & $0.744\pm 0.023$ & $0.937\pm 0.011$ \\
VGGNet-19      & $0.750\pm 0.013$  & $0.709\pm 0.023$  & $0.776\pm 0.005$ & $0.748\pm 0.013$ & $0.756\pm 0.010$ & $0.946\pm 0.004$ \\ \hline
\textbf{AVG}   & $0.764\pm 0.023$  & $0.718\pm 0.022$  & $0.779\pm 0.018$ & $0.764\pm 0.023$ & $0.766\pm 0.021$ & $0.948\pm 0.008$ \\ \hline
\end{tabular}
\label{tab:results-feat}
\end{table}

As we can note from both tables, there is a notable improvement for all metrics from scenario 1 to scenario 2. In general, the clinical features impacted positively in all models. In terms of balanced accuracy, the average model was improved in almost 7\%. The overall improvement considering all metrics is also around 7\%. Although the improvement is notable, we also applied the statistical test for this experiment. As expected, the test shows that all models from scenario 2 are statistical different than the ones from scenario 1. Therefore, based on the test and the results present in the tables, we conclude that for this experiment, the best option is including the clinical features.

\begin{figure}
    \centering
  \subfigure[Scenario 1\label{fig:conf_mats-a}]{
      \includegraphics[width=0.45\linewidth]{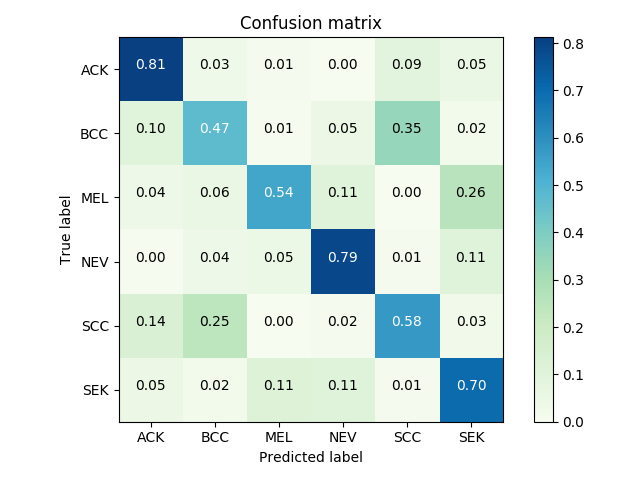}}
    \quad
    \subfigure[Scenario 2\label{fig:conf_mats-b}]{%
        \includegraphics[width=0.45\linewidth]{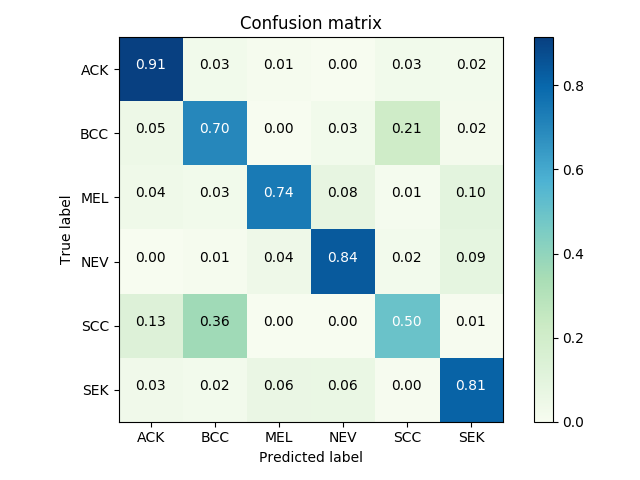}}

    \subfigure[Scenario 1\label{fig:roc-a}]{
      \includegraphics[width=0.45\linewidth]{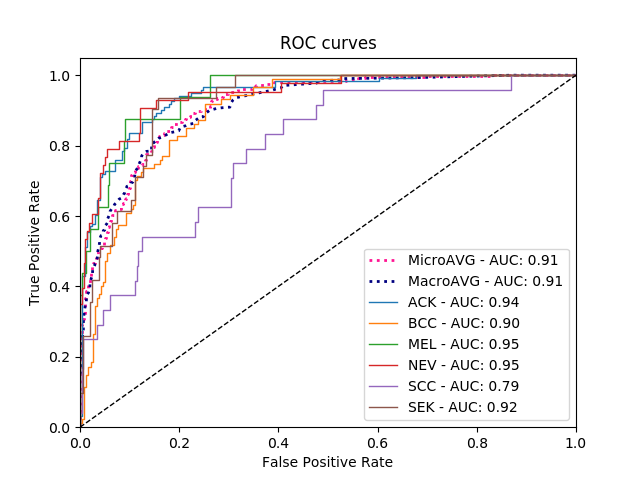}}
    \quad
    \subfigure[Scenario 2\label{fig:roc-b}]{%
        \includegraphics[width=0.45\linewidth]{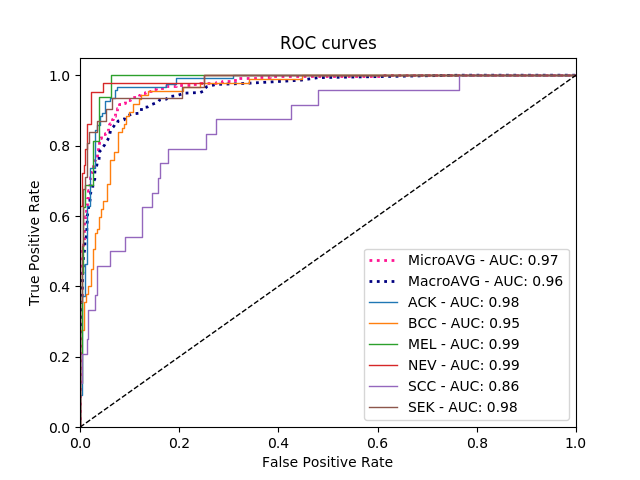}}

    % \subfigure[Scenario 1: MobileNet\label{fig:conf_mats-e}]{
    %   \includegraphics[width=0.4\linewidth]{src/img/graphs/mobilenet-no-feat.png}}
    % \quad
    % \subfigure[Scenario 2: MobileNet\label{fig:conf_mats-f}]{%
    %     \includegraphics[width=0.4\linewidth]{src/img/graphs/mobilenet-feat.png}}
        
  \caption{The confusion matrices (above) and ROC curves (below) for ResNet-50 considering for both scenarios}
  \label{fig:conf_mats} 
\end{figure}

Observing only the results for scenario 2 in Table \ref{tab:results-feat}, in terms of balanced accuracy, we note all models with almost the same performance, except ResNet-50, which is around 4\% above compared to the rest of the models. In order to better investigate the influence of the clinical feature in this model, in Figure \ref{fig:conf_mats} is depicted the ResNet-50 confusion matrix and ROC curve for each scenario \footnote{As we would have 12 confusion matrices and 12 ROC curves, we decided to present a thorough analysis only for the best model. However, the remaining results are quite similar.}. In general, it is possible to note an improvement for all labels. However, the model is still confusing SCC and BCC quite often. This results is in accordance with the analysis provided in section \ref{sec:data_analysis}, in which we show that both lesions share almost the same value of features. In fact, even dermatologists get confusing regarding these two lesions. It is very challenging to differentiate them even using a dermatocospe since they are very similar, as shown in Figures \ref{fig:samples-bcc} and \ref{fig:samples-scc}. Nonetheless, confusing SCC and BCC is not quite a problem since both are skin cancer and need to be removed and sent to biopsy. The real problem is confusing them with ACK, which is just a minor skin disease that is treated without a surgical process, for example. For this lesion, the model does a fair job, even though there is still room for improvement.

\begin{figure}
    \centering
  \subfigure{
      \includegraphics[width=0.3\linewidth]{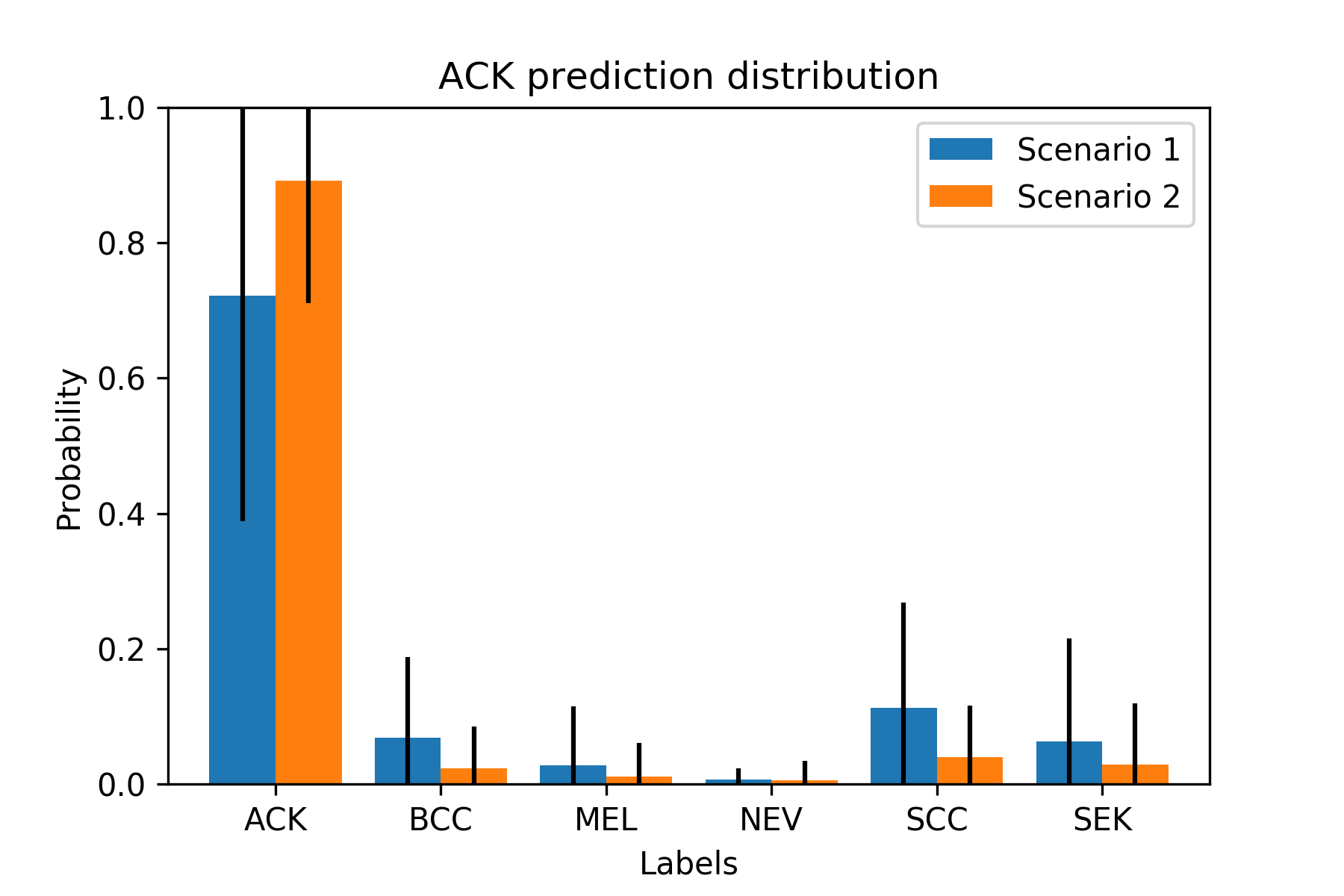}}
    \quad
  \subfigure{%
      \includegraphics[width=0.3\linewidth]{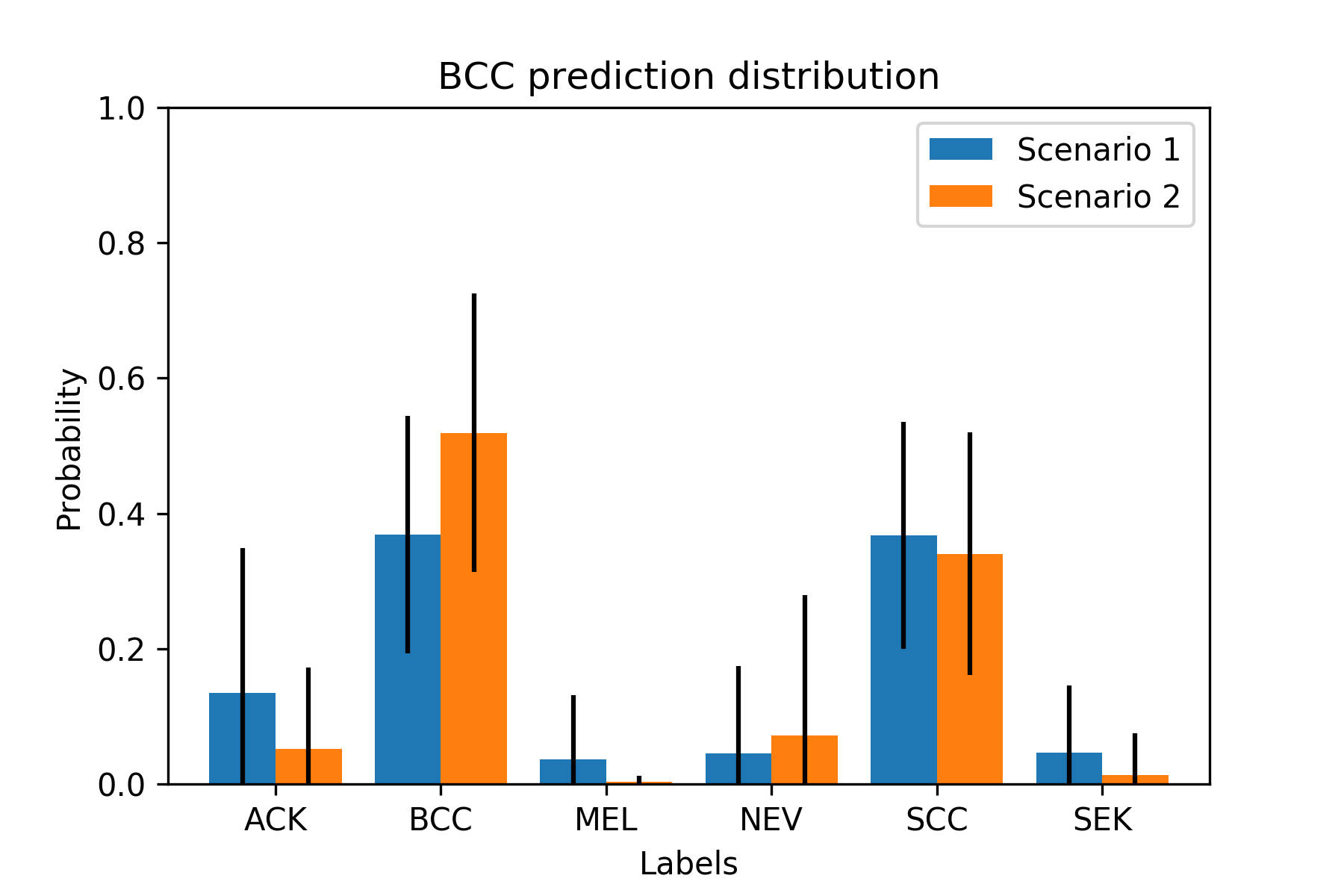}}
    \quad
  \subfigure{%
      \includegraphics[width=0.3\linewidth]{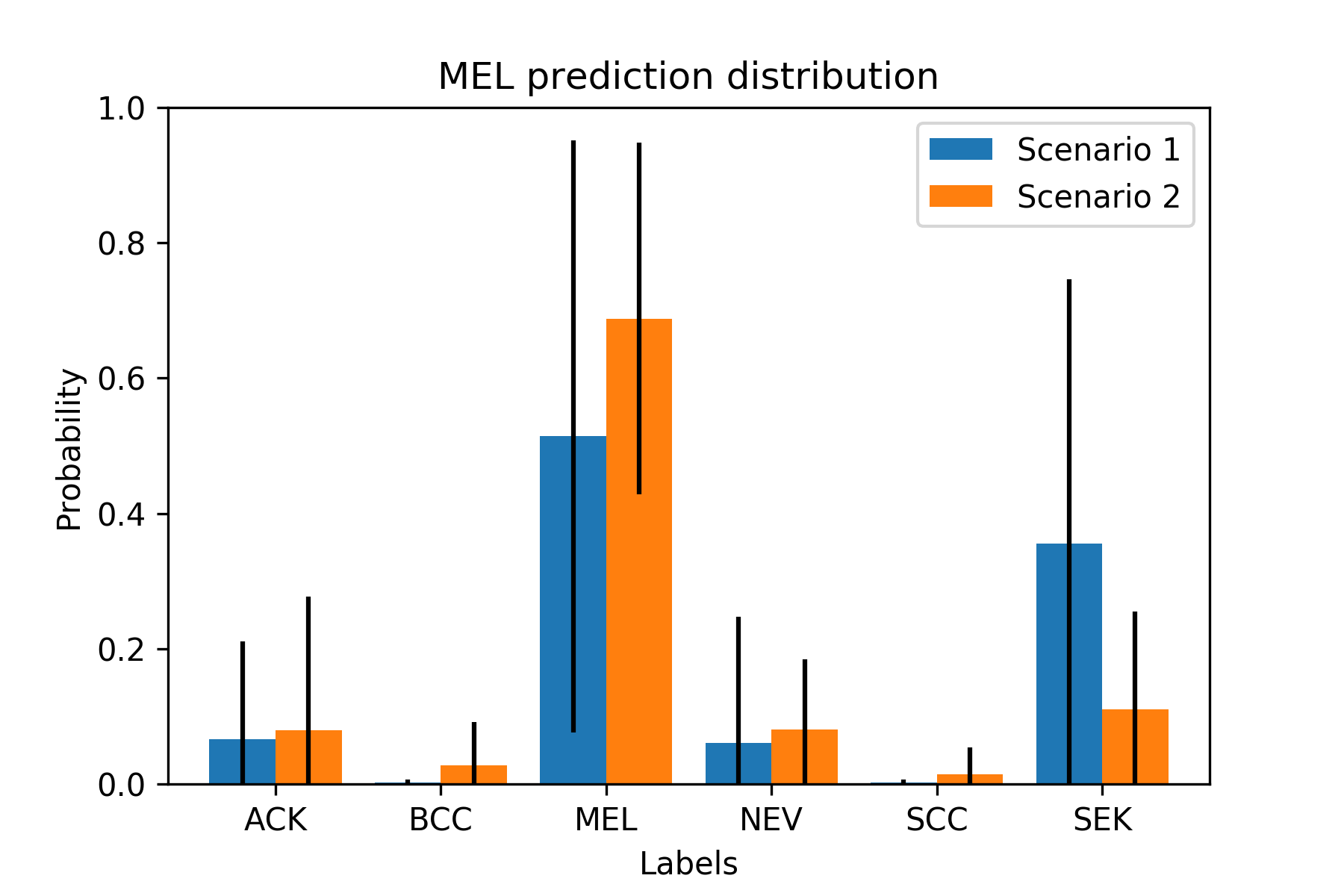}}

    \subfigure{
      \includegraphics[width=0.3\linewidth]{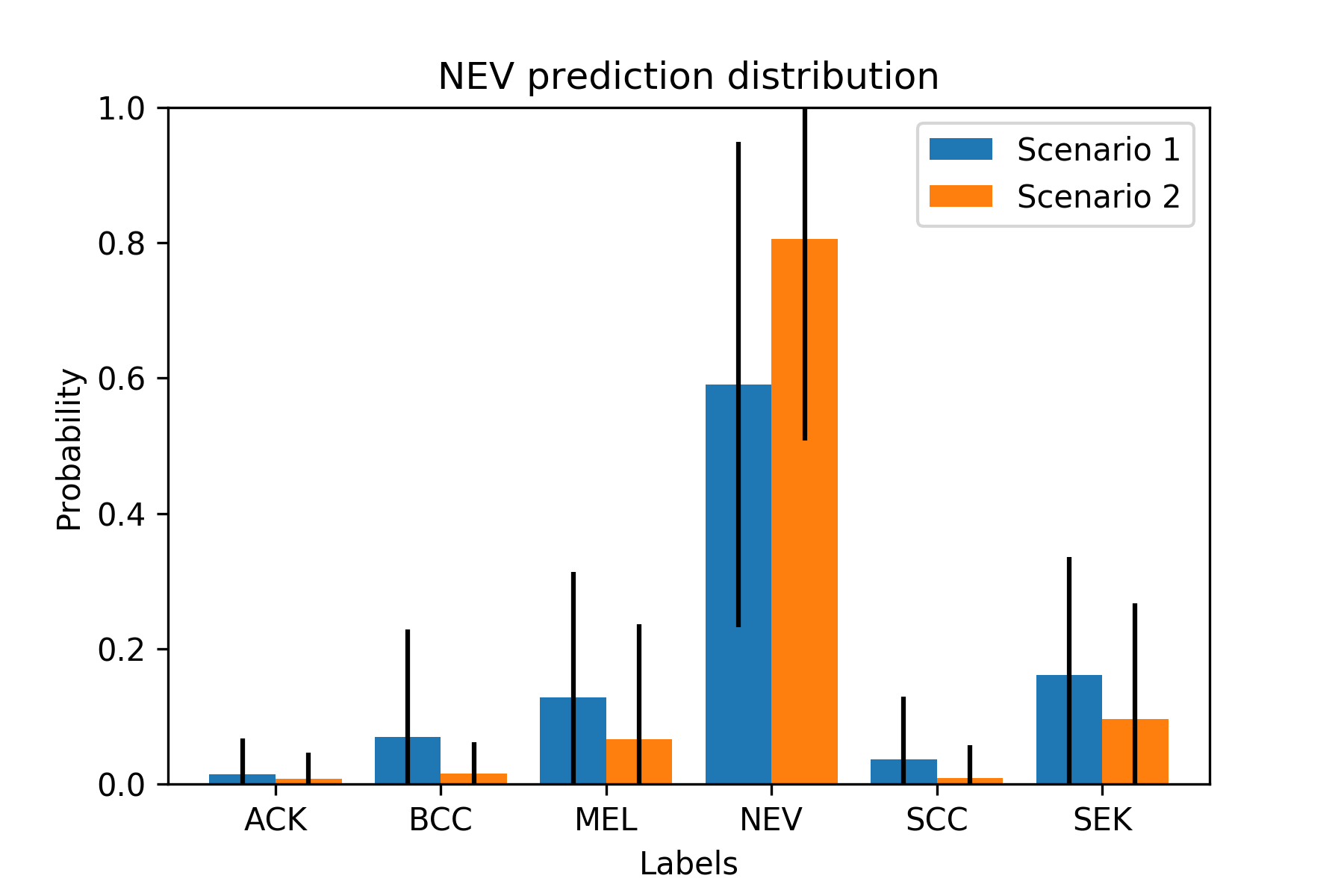}}
    \quad
  \subfigure{%
      \includegraphics[width=0.3\linewidth]{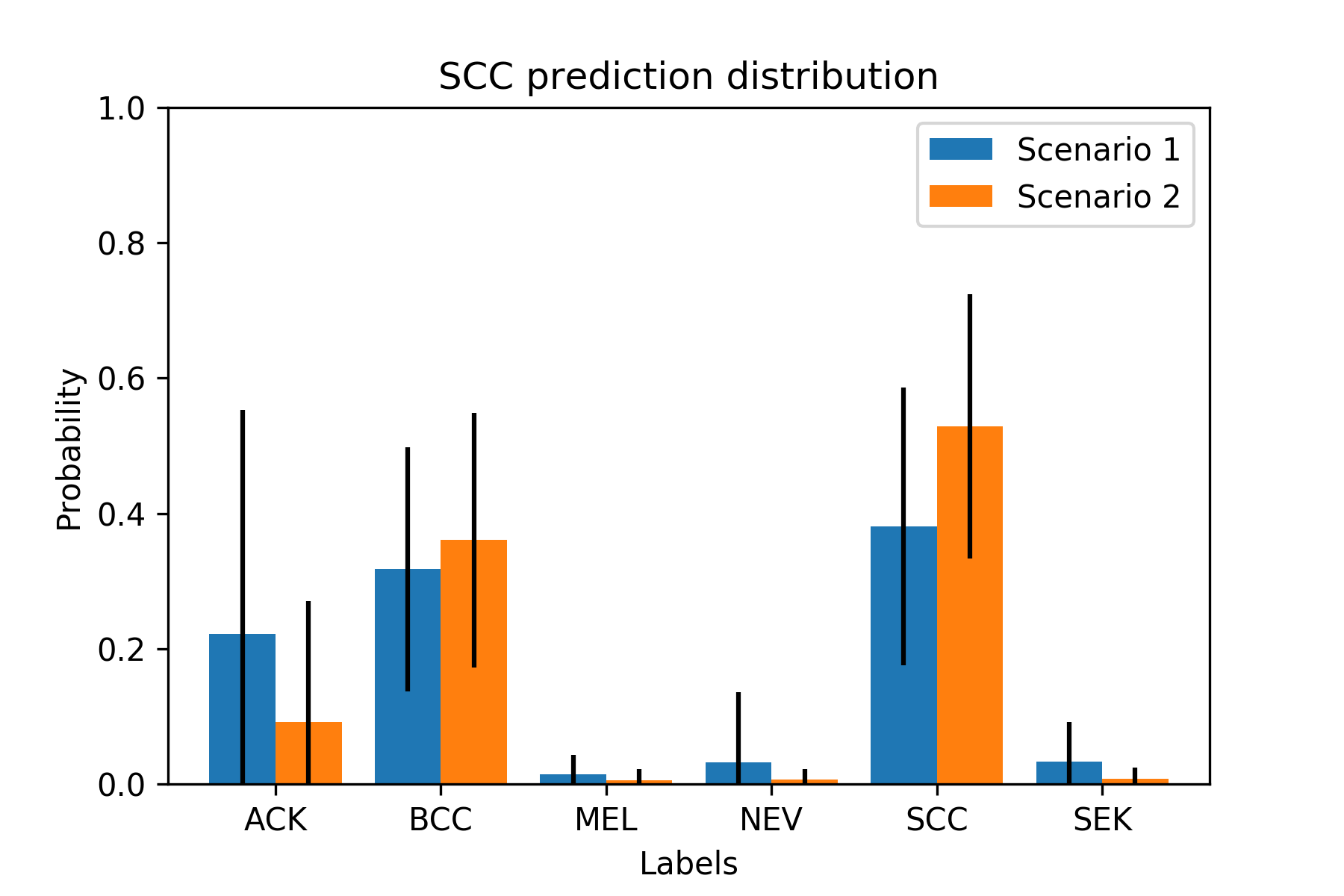}}
    \quad
  \subfigure{%
      \includegraphics[width=0.3\linewidth]{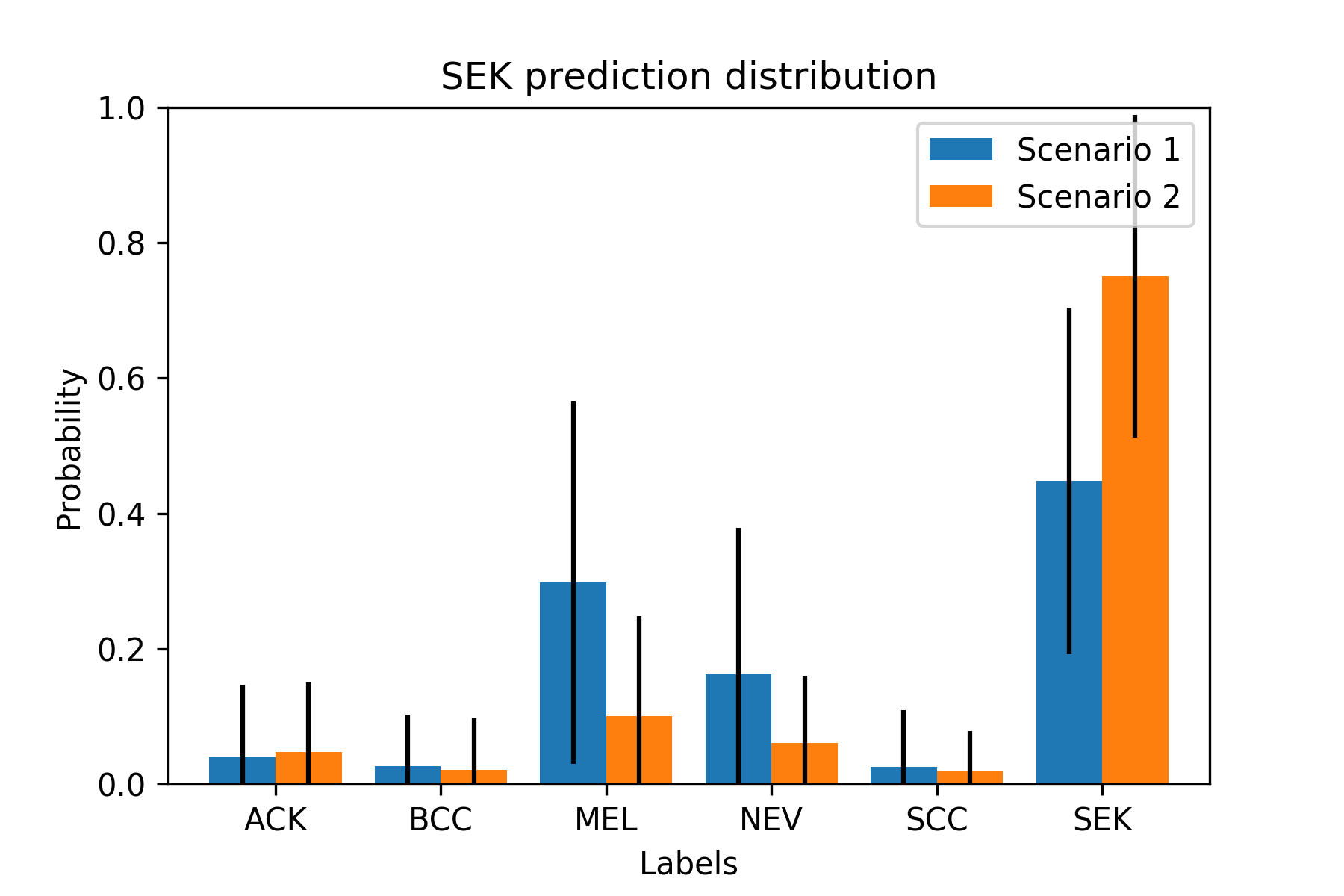}}

  \caption{The probability distribution generated by the ResNet-50 for each lesion in both scenarios}
  \label{fig:dists} 
\end{figure}

To conclude this experiment, we analyze the effect of the clinical features in the probability distribution generated by the ResNet-50 model. In Figure \ref{fig:dists} is depicted the distributions obtained by the same model for each lesion in both scenarios. As we may see, the distributions for ACK, MEL, NEV and SEK are improved from scenario 1 to scenario 2 by increasing the own label probability and decrease the remaining ones. On the other hand, the distributions for SCC and BCC are almost the same, which is in accordance with the previous results. In overall, the results presented in this section confirm the hypothesis raised by Briker \textit{et al.} \cite{brinker2018} that clinical features is an important piece of information to be used in deep learning models in order to improve the skin cancer detection. Nonetheless, we show it is not effective for all kind of lesions, which is the case of SCC and BCC detection.

\section{Conclusion}
In this paper, we presented a study to analyze the impact of the patient clinical information on the skin cancer detection using deep learning models. First, we introduced a new dataset containing clinical images, acquired from smartphones cameras, and patients clinical information. Next, we presented a straightforward approach to combine the image and clinical features using convolutional neural networks. We implemented this approach for different CNN models and applied them for the presented dataset. The results indicated that the clinical features provided a substantial improvement for all investigated models in this work. It was possible to note that the clinical features are important pieces of information that may help to overcome the lack of large amount of data. Nonetheless, the clinical features used are not helpful for all kind of lesions. As we discussed, they were not able to improve the classification of SCC/BCC lesions, since its features are quite similar. In general, this work showed the importance of clinical features in skin cancer detection and confirms the hypothesis that patient clinical information is helpful for this task. As a future work, we intend to improve the aggregation approach and include a hierarchical classification in order to improve the SCC and BCC detection. In addition, we are already working on the inclusion of more clinical features and more images in our dataset.

\section*{Acknowledgments}
This study was financed in part by the Coordena\c{c}\~{a}o de Aperfei\c{c}oamento de Pessoal de N\'{i}vel Superior - Brasil (CAPES) - Finance Code 001; the Conselho Nacional de Desenvolvimento Cient\'{i}fico e Tecn\'{o}logico (CNPq) - grant n.309729/2018-1 - and the Funda\c{c}\~{a}o de Amparo \'{A} Pesquisa e Inova\c{c}\~{a}o do Esp\'{i}rito Santo (FAPES) - grant n. 575/2018. We also thank all the members of the Dermatological Assistance Program (PAD-UFES) and the support of NVIDIA Corporation with the donation of a Titan X GPUs used for this research.

\bibliographystyle{unsrt}
\bibliography{references}

\end{document}